\documentclass[useAMS,usenatbib]{mn2e}
\usepackage{graphicx}

\newcommand\bb[1] {   \mbox{\boldmath{$#1$}}  }
\newcommand\del{\bb{\nabla}}

\newcommand\btimes{\bb{\times}}

\bibpunct{(}{)}{;}{a}{}{,}

\title[Accumulation of solid bodies in turbulent discs]{On the accumulation of solid bodies in global turbulent protoplanetary disc models}

\author[S.Fromang and R.P.Nelson]{ S\'ebastien
  Fromang$^{1}$\thanks{E-mail:s.fromang@qmul.ac.uk} \&  Richard
  P. Nelson \\
$^{1}$ Astronomy Unit, Queen Mary, University of London, 
Mile End Road, London E1 4NS}

\begin{document}

\date{Accepted; Received; in original form;}

\maketitle

\label{firstpage}

\begin{abstract}
We study the migration of solid bodies in turbulent
protoplanetary accretion discs by means of global MHD simulations.
The bodies range in size from 5 centimetres up to 1 metre, and so include 
objects whose migration is expected to be the most rapid
due to gas drag interaction with the disc. As they drift inward through the
disc, some of them are trapped in regions where gas pressure maxima are
created by long lived anticyclonic vortices. This accumulation is very 
efficient, locally increasing the dust--to--gas ratio by a factor $ > 100$ 
in some cases. We discuss the possible implications of this result for 
theories of planet formation.
\end{abstract}


\section{Introduction}

A crucial element of any planet formation theory is understanding
how the dust component of protoplanetary discs builds up to form larger
objects that eventually become planetesimals and planets. This has obvious
relevance to the formation of rocky terrestrial planets and asteroids,
and according to the core--accretion model of gas giant formation
(e.g. Pollack et al. 1996), is a key stage in the formation of giant planets.

The growth and evolution of solid bodies of all sizes is therefore
an important topic to study. A key issue relates to the growth and
survivability of metre--sized objects, which are expected to undergo
rapid inward migration due to gas drag \citep{weidenschilling77}. Such bodies
are predicted to migrate into the central star within $\sim 100$ years of 
formation, raising questions about how planetary systems form at all.
Models of planet formation in smooth, laminar protoplanetary discs,
beginning with small dust grains that must
grow through this size range, require a radially extended reservoir 
of solid matter to overcome this problem (e.g. \citeauthor{weidenschilling&cuzzi93} 
\citeyear{weidenschilling&cuzzi93}).

In this letter, we investigate the effect MHD turbulence resulting
from the magneto--rotational instability (MRI) in the disc might have on the
dynamics of such objects. It seems likely
that the MRI is the source of turbulence which transports angular
momentum in protoplanetary discs. The velocity and density fluctuations
that result will modify the drag force exerted on solid bodies and
will affect their dynamics. These fluctuations can range from
being relatively small variations, up to being 
substantial modifications of the disc
radial profile caused by radial variations in the turbulent stresses
(e.g. \citeauthor{hawley01} \citeyear{hawley01}; \citeauthor{steinacker&pap02} 
\citeyear{steinacker&pap02}).

Because of the global nature of this problem, in which solid bodies
will drift through a substantial fraction of the
disc radius, we set up global protoplanetary disc
simulations. As in \citet{pap&nelson03a}, we have developed
cylindrical discs models to limit the large computational cost
of these calculations. We note that there has been a recent study of this
problem by \citet*{johansenetal05}, who find transient trapping
of solid bodies in density maxima generated by MHD turbulence.
The approach used in
their study, however, was local rather than the global approach
we have taken.

The plan of the paper is as follows. In section~$2$, we describe our
numerical schemes,
and in section~$3$ we describe our results. We discuss their consequences
and limitations in section~$4$. 

\section{Simulations}

\subsection{Algorithms}

Two similar MHD Eulerian codes were used to compute the simulations 
presented in this paper: GLOBAL \citep{hawley&stone95} and NIRVANA 
\citep{ziegler&yorke97}. Both use finite difference techniques combined with 
the Constrained Transport algorithm to evolve the MHD
equations. They differ, however, in their treatment of the solid
phase: GLOBAL treats the solid bodies as a second fluid; NIRVANA
treats the solid bodies as particles. This reflects the sizes of the
objects considered in each of our simulations.

\subsubsection{The two fluid approach}

GLOBAL was extended to describe the dust particle evolution as a 
second, pressureless fluid. This second fluid interacts with the gas via 
the drag force exerted by the latter. 
The drag force $\bb{F_{drag}}$ takes the form
\begin{equation}
\bb{F_{drag}}=\frac{m_p}{\tau_s}(\bb{v_{g}}-\bb{v_{d}}) \, .
\end{equation}

\noindent
Here $m_p$ is the mass of the particle, ($\bb{v_{g}} - \bb{v_{d}}$) 
is the gas velocity relative to the solid component, and $\tau_s$ is the dust 
stopping time. The small dust particles examined using GLOBAL are in 
the Epstein regime, so $\tau_s$ is defined by
\begin{equation}
\tau_s=\frac{\rho_{s} a}{\rho c_s} \, ,
\end{equation}

\noindent
where $\rho$ is the gas density, $c_s$ the speed of sound, $\rho_s$ the
density of the dust particles and $a$ their size. Because 
of the short stopping time of the small particles studied in this work, the 
effect of the drag force is computed implicitly.

\subsubsection{The N--body approach}
The two fluid approximation breaks down for
large solid bodies, so we adapted NIRVANA to include
coupling between the fluid and individual solid objects
represented by particles. We consider metre-sized bodies,
whose drag force interaction is given by Stokes Law for the disc parameters
we have adopted. The stopping time, $\tau_s$, then becomes 
(e.g. \citeauthor{weidenschilling77} \citeyear{weidenschilling77})
\begin{equation}
\tau_s = \frac{2 \rho_s a^2}{9 \eta}
\label{drag-particles}
\end{equation}
where $\eta$ is the gas viscosity defined by 
\begin{equation} 
\eta \simeq \frac{\lambda c_s}{3} \rho
\label{viscosity}
\end{equation}
and the mean free path of molecules 
$\lambda = (n_{{\rm H}_2} \sigma_{{\rm H}_2} )^{-1}$.
We adopt a value of $\sigma_{{\rm H}_2} =10^{-15}$ cm$^2$ 
for the collision cross section of molecular hydrogen, and assume that
$n_{{\rm H}_2} = \rho/m_{{\rm H}_2}$, where $m_{{\rm H}_2}$ is the mass 
of a hydrogen molecule (e.g. \citeauthor{rafikov04} \citeyear{rafikov04}).
We employ linear interpolation, using information
from the surrounding eight cells, to obtain relevant physical quantities at the
location of the particles.

In one simulation performed (model N1), the particles interacted with
the disc through the drag force only. A second simulation
(model N2) was performed in which the particles also experienced
gravitational interaction with the disc. In this case the 
axisymmetric component of disc self-gravity was included to
ensure that the velocity difference between solid particles
and the gas was computed correctly (see \citeauthor*{fromangetal05} 
\citeyear{fromangetal05} for
a description of the implementation). Simulations of low mass planets
in turbulent discs indicate that the gravitational interaction between
embedded objects and the turbulent fluctuations can have important consequences
for the orbital evolution \citep{nelson&pap04b,nelson05}. 
It is for this reason we computed run N2 to examine the effect this may have
on the evolution of smaller bodies that also interact with the disc gas 
through the drag force.

\subsection{Model parameters}

\begin{table}
\begin{center}
\begin{tabular}{@{}lccccc}
\hline\hline
Model & dust & Number of  & Resolution & Particle size \\
 & description & dust particles &  &  (in cm) \\
\hline\hline
G1 & fluid & No particle & $(260,152,44)$ & 5 \\
G2 & fluid & No particle & $(260,152,44)$ & 25 \\
\hline
N1 & N--body & $3000$ & $(260,604,44)$ & 100 \\
N2 & N--body & $500$ & $(260,604,44)$ & 100 \\
\hline\hline
\end{tabular}
\caption{Model parameters of the runs presented here: Column~$1$ gives the 
name of the model. The first two (G1 and G2) are performed with GLOBAL and 
the last two (N1 and N2) with NIRVANA. The algorithm used to evolve the 
dust component (fluid vs N--body) is described in column~$2$. For the particle 
approach, column~$3$ gives the number of dust particles used. Finally, 
column~$4$ gives the resolution of the run and column~$5$ the size of the 
solid bodies, in centimetres.}
\label{model properties}
\end{center}
\end{table}

The parameters of the models that we computed here are described in 
table~\ref{model properties}. We performed four models, labelled G1, G2, 
N1 and N2. They correspond to increasing particle sizes: $5$ cm, $25$ cm, 
and $1$ m. The first two, for which there is good coupling between 
the gas and the dust, were computed with GLOBAL using the two fluid approach. 
In models N1 and N2 the large particles start to 
decouple from the gas, so the N--body approach implemented in NIRVANA was used.
In the following, we describe the disc model parameters 
and the properties of the MHD turbulence that 
develops as a consequence of the MRI. We then describe the simulation results.

\section{Results}

\subsection{Disc model}
\label{Discmodel}

\begin{figure}
\begin{center}
\includegraphics[scale=0.34]{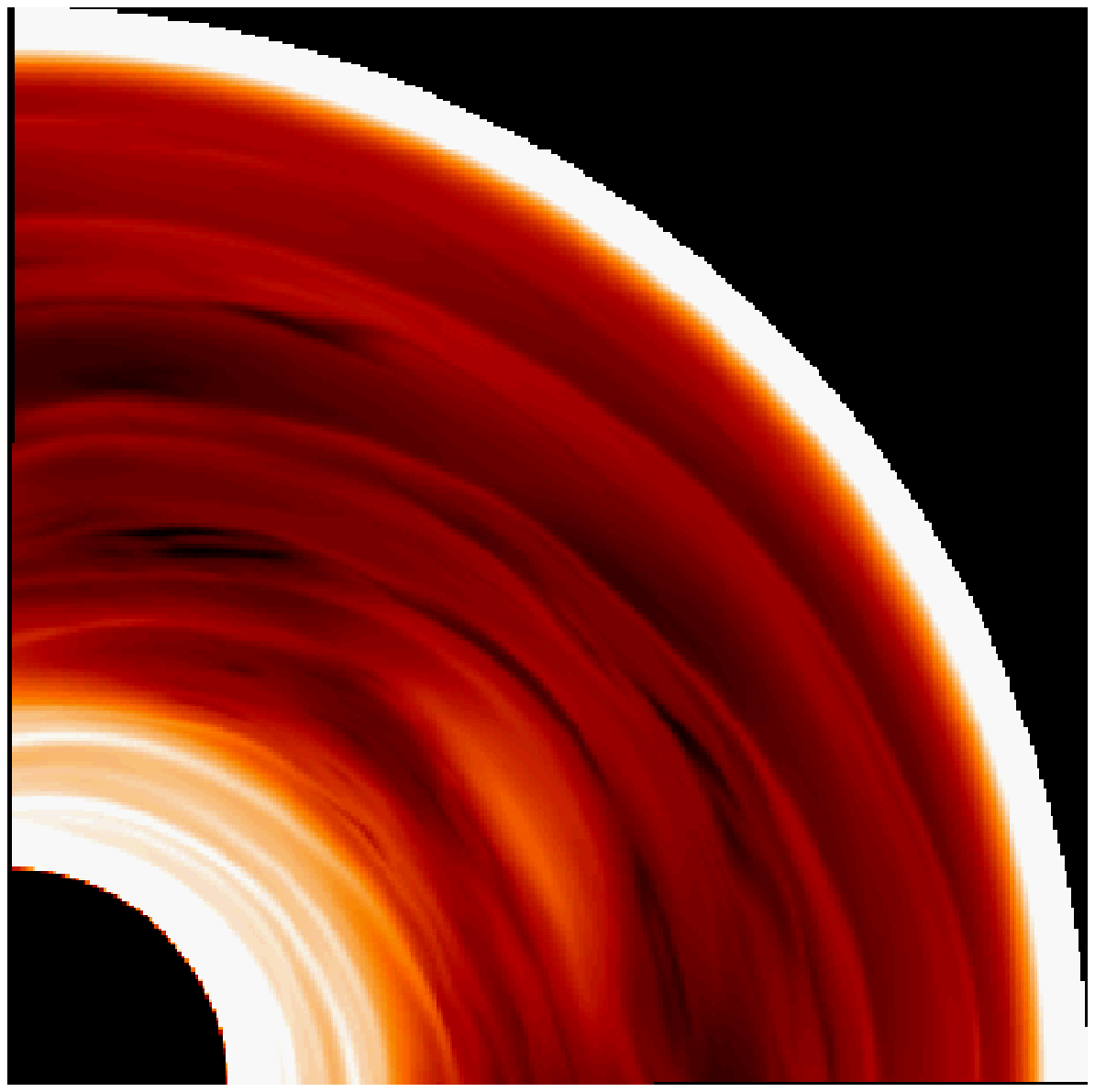}
\includegraphics[scale=0.34]{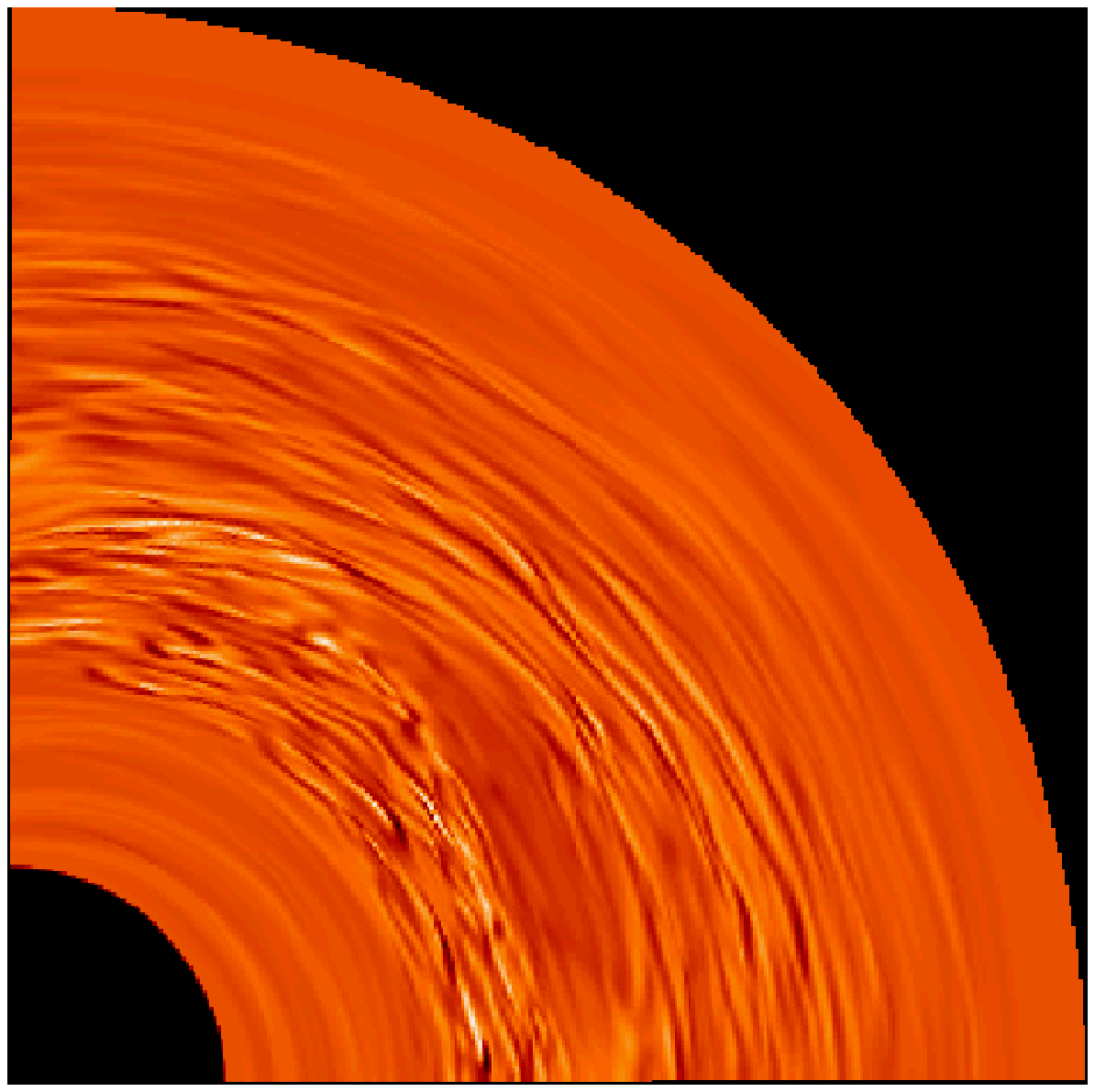}
\caption{Logarithm of the density in the disc ({\it left panel}) and 
vorticity in the local rotating frame ({\it right panel}). Both 
quantities are calculated at the end of model G1. A region of negative 
vorticity 
is seen in the middle of the disc (at a location $r \sim 2.5$ and 
$\phi \sim \pi/6$), well correlated with a density increase on the left panel.}
\label{log_rho}
\end{center}
\end{figure}

The turbulent disc structure is computed following the procedure described 
in \citet{nelson05}: $\rho$ varies like $r^{-1}$ and 
the equation of state is locally isothermal, with $c_s \propto r^{-0.5}$. 
The computational domain ranges in radius 
from $R_{min}=1$ 
to $R_{max}=5$ and its vertical extend is $\Delta H=0.28$. In order to
reduce the computational cost of models G1 and G2, the azimuthal domain 
only covers the interval $[0,\pi/2]$. In that case, the resolution 
is $(N_r,N_{\phi},N_z)=(260,152,44)$. In models N1 and N2, which 
cover the full azimuthal domain, the number of grid cells in $\phi$
is four times larger. 

When converting from computational to physical units, 
we assume that $r=2.3$ corresponds to 5 AU. The disc mass and
density are scaled such that the total disc mass would be 0.06 M$_{\odot}$
within 40 AU (i.e. approximately three times the minimum mass nebula).
The orbital period at the disc inner edge thus corresponds to $\simeq 6$ years.
We assume that the internal density of solid bodies $\rho_s = 3$ g cm$^{-3}$.

In all of our models, turbulence is initiated in the
disc with a non--zero net flux toroidal magnetic field. When the MRI
starts to saturate, the field is scaled down by a factor of two, while
the density profile is reset to its initial value \citep{nelson05}. 
The simulation is then restarted. This procedure gives a typical 
volume--averaged $\alpha$ value computed from the Maxwell and Reynolds 
stress of $\sim 10^{-2}$. In both GLOBAL and NIRVANA, a quasi--steady state is 
obtained at $t \simeq 300$ (measured throughout this paper in orbits at 
the inner edge of the disc), when
the solid component is introduced. The turbulent state of the disc
maintains  a rough statistical steady state during the run time 
($\sim$ 180--200 orbits). The disc structure is illustrated by 
figure~\ref{log_rho}, where the left panel shows the density in the 
midplane of the disc at the end of model G1. A local increase in the 
density is visible in the middle of the disc. To better understand its nature, 
we plot in the right panel of figure~\ref{log_rho} the vorticity $\omega$ which 
is calculated at each radius in the local rotating frame according to:

\begin{equation}
\omega=\del \btimes \bb{\bar{v_g}} \, .
\end{equation}

\noindent
Here $\bb{\bar{v_g}}$ is the difference between the local gas velocity and the 
azimuthally averaged velocity. White areas correspond to large 
positive values of $\omega$ and dark zones to large negative values. A region 
with a lower value of $\omega$ than the background of the disc is seen, which
is well 
correlated with the density increase shown on the left panel. This is the 
signature of an anticyclonic vortex \citep{Johansen&klahr05}. Note that the 
decrease in $\omega$ that is created is of smaller amplitude than the vorticity 
fluctuations that result from the MHD turbulence. Similar vortices are observed
in runs N1 and N2, and we observe these
structures to survive for the duration of the simulations. Their 
nature and formation 
mechanism are discussed further and compared with previous work in 
section~\ref{discussion}.

\subsection{Models G1 and G2}
\label{G1G2}

\begin{figure}
\begin{center}
\includegraphics[scale=0.34]{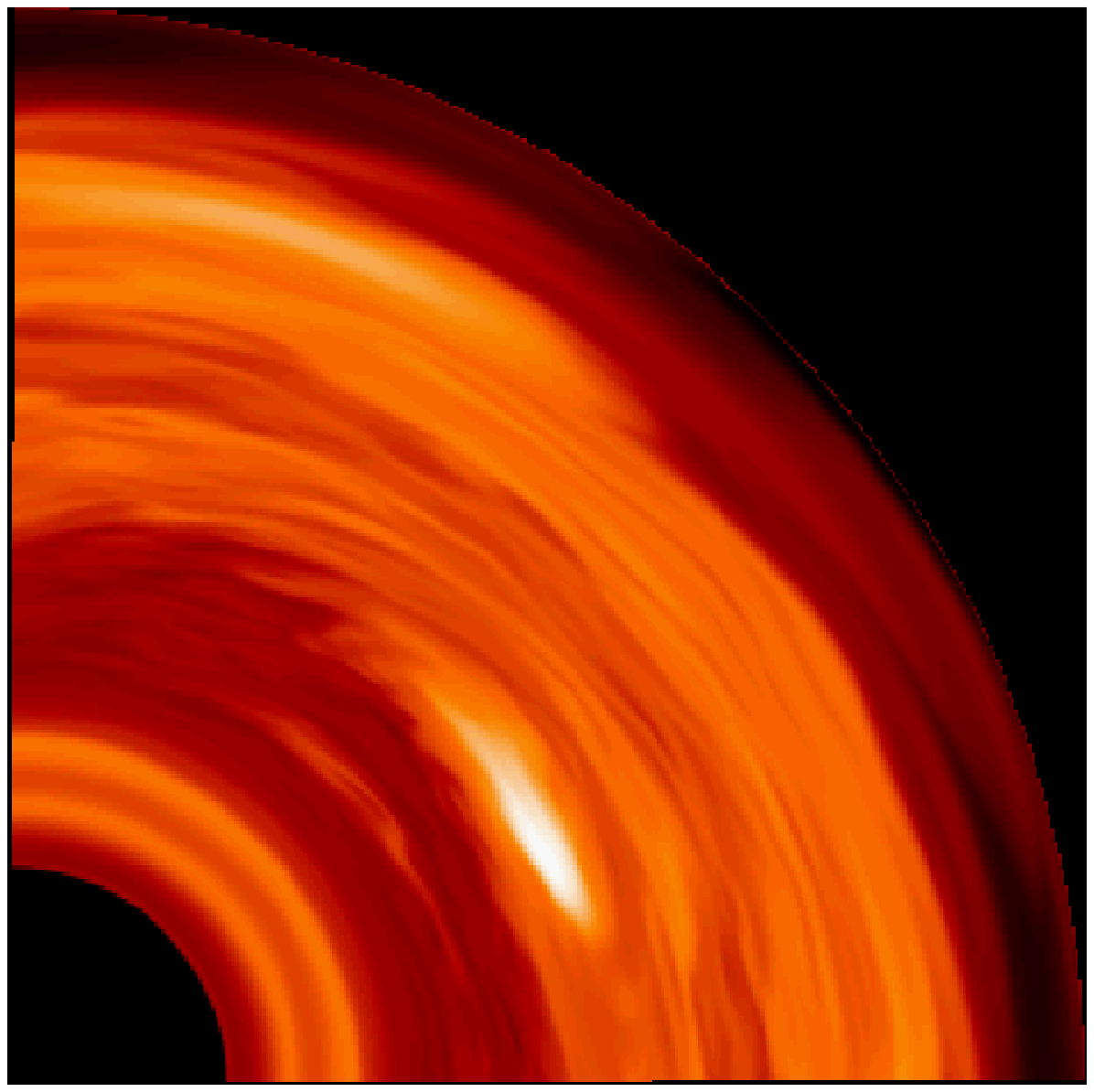}
\includegraphics[scale=0.34]{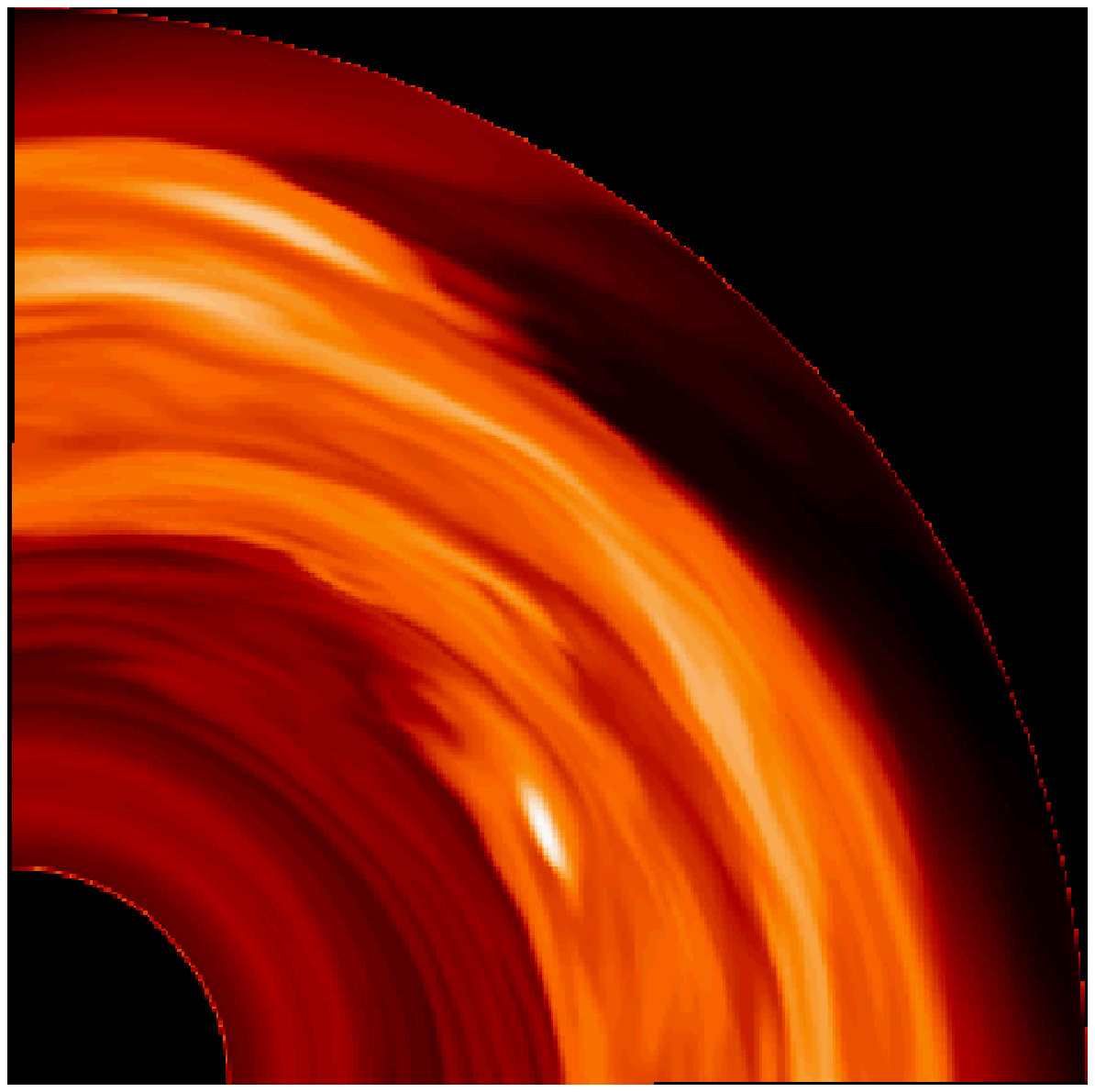}
\caption{Logarithm of the dust--to--gas ratio at the end of model G1 ({\it left 
panel}) and G2 ({\it right panel}). Its maximum value respectively reaches 
$0.34$ and $1.41$.}
\label{log_dust}
\end{center}
\end{figure}

At time $t=300$, the dust is introduced such that $\rho/\rho_d=100$ 
everywhere in the disc and the evolution is followed until
$t=477$. The parameter $\Omega \tau_s$ (where $\Omega$ is the Keplerian angular 
velocity) is independent of radius in these models.
Two cases are investigated. Model G1 corresponds to 
$\Omega \tau_s=0.1$, and  model G2 has
$\Omega \tau_s=0.5$. Given the model normalization 
described above, 
these parameters respectively correspond to solid bodies whose diameter
is equal to $5$ and $25$ centimetres.

During the simulations, the dust drifts radially toward the
central object and the spatial distribution evolves. The logarithm
of the gas--to--dust ratio in the $(r-\phi)$ plane is shown in figure 
\ref{log_dust} for 
model G1 ({\it left panel}) and G2 ({\it right panel}) at time $t=176$. The 
striking feature of both of the snapshots is the accumulation of 
dust in a region of the disc close to $r=2.7$. The peak value of the dust to 
gas ratio at that time is $0.34$ and $1.41$ for model G1 and G2,
which indicates an increase of respectively $34$ and $141$ during the
course of the simulation. A comparison between 
figures~\ref{log_rho} and \ref{log_dust} shows that this accumulation occurs 
at a location corresponding to a local increase of the gas density
(or, equivalently, of the gas pressure). We have shown in  
section~\ref{Discmodel} that this pressure extremum 
corresponds to an anticyclonic vortex. The 
fact that solid bodies are efficiently trapped in such a vortex has been 
previously reported in the literature on several occasions 
\citep{barge&sommeria95,godon&livio00,johansenetal04}. The novel result of our 
work is that this long--lived vortex is self-consistently generated by 
the MHD turbulence itself.

\begin{figure}
\begin{center}
\includegraphics[scale=0.45]{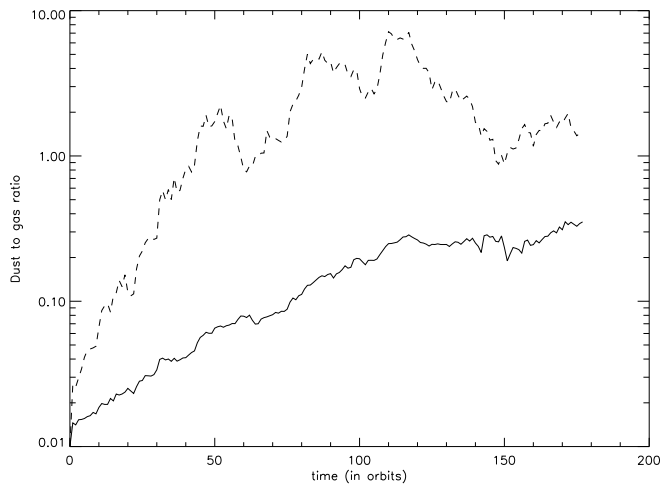}
\caption{Time history of the maximum value reached by the gas to dust
  ratio for model G1 ({\it solid line}) and G2 ({\it dashed line}).}
\label{gas2dust ratio}
\end{center}
\end{figure}

Finally, figure~\ref{gas2dust ratio} plots the time history of the
maximum value reached on the grid by the gas--to--dust ratio for
models G1 and G2. Both curves 
show an increase with time. However, there is a clear
tendency for larger particles to accumulate more quickly in gas
pressure maxima, as they undergo faster radial migration,
and the maximum value for the dust--to--gas ratio 
reached in model G2 is greater than unity.
In both models, the value of the dust--to--gas ratio
saturates after about $120$ orbits in model G1 and after $50$ orbits
in model G2. This is because the outer regions of the disc
are depleted at that time and there is no more solid material to
supply the region where dust accumulates. 

\subsection{Models N1 and N2}
\begin{figure}
\begin{center}
\includegraphics[scale=0.5]{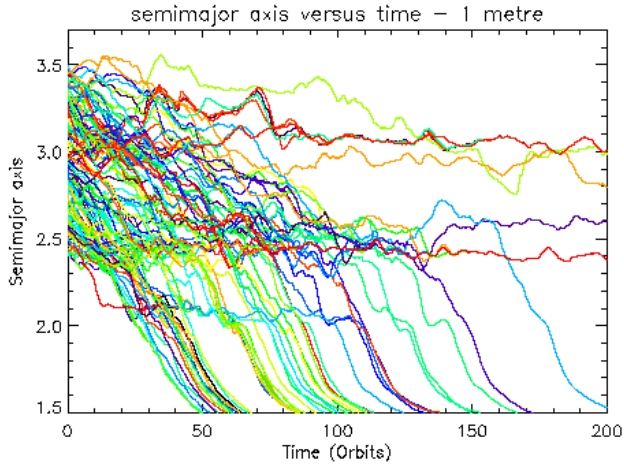}
\caption{This figure shows the radial trajectories of 120 representative
solid bodies from simulation N1. Note that there are two regions in the
disc where particles are trapped, corresponding to local
pressure maxima in the disc. Very similar results were obtained
in simulation N2.}
\label{N1fig}
\end{center}
\end{figure}
Simulations N1 and N2 employed 3000 and 500
particles, respectively,  to trace the trajectories
of 1 metre sized solid bodies. 
Their initial radial and azimuthal locations
were distributed randomly in a narrow annulus between $2.4 \le r_p \le 3.2$,
and they were initiated with Keplerian circular velocities.
For the disc model we adopt, the stopping time for 
simulations N1 and N2 are given by
$\Omega \tau_s=1.15 \left(\frac{r_p}{2.5}\right)^{-1}$. Between radii
1.5 -- 3.2,  $\Omega \tau_s$ thus ranges between $\sim$ 1.92 and 0.85.
The radial trajectories of 120 particles from simulation N1 are
shown in figure~\ref{N1fig}. The simulation ran for 200 orbits
at the disc inner edge. The inward drift of particles due to gas drag
is apparent. For those particles that undergo global migration 
through the disc, we find reasonable overall agreement 
in the migration time compared with that
obtained in an equivalent laminar disc simulation.
It is clear, however, that there are particles
that do not undergo substantial inward drift over the simulation run time,
but instead become trapped at specific radii. Figure~\ref{N1fig} shows that
there are four separate radii around which particles become
concentrated.
Most of these particles become trapped because they become concentrated
in anticyclonic vortices that form in the turbulent flow, and maintain
their identity for the duration of the simulation.  For a given vortex,
the trapped particles eventually find themselves being concentrated into a
single grid cell, such that they then maintain very similar trajectories
for the duration of the simulation.

Out of 3000 particles considered, 1356 migrate interior to $r=1.5$ after
200 orbits, without becoming trapped at intermediate radii. 
There are three distinct vortices that form in simulation N1 that lead to 
particles being concentrated. One is located at radius $r \simeq 2.4$, and traps
1276 particles. Another vortex forms at radius $r \simeq 2.6$, and contains
24 particles.
This vortex is observed to be weaker than the former one,
explaining the reduced particle concentration.
At the end of the simulation there are 3 particles orbiting
at radii $r \simeq 2.8$, but these are not contained within an obvious vortex,
but appear to be tail-enders whose inward drift has been disrupted
through turbulence--induced perturbations. A third vortex orbits at 
$r \simeq 3.0$, and contains 341 particles at the end of the simulation.
Statistically, we find that
just under 50 percent of the particles migrate substantially
through the disc, and just over 50 percent become concentrated 
within the vortices that form.

A similar picture emerges from run N2, which included the gravitational
interaction between solid bodies and the disc. This simulation resulted
in particles becoming concentrated in vortices at radii 
$r \simeq 2.5$ and $r \simeq 2.9$.
The degree of concentration was similar to that in run N1,
with 40 percent of particles
migrating interior to $r=1.5$ and 60 percent being concentrated
further out in the disc.  It is clear that
gas drag dominates the evolution for 1 metre-sized bodies, with 
perturbations to their motion induced by
disc gravity having little effect. Simulations that include larger
bodies ($a \ge 10$ metres) indicate that the concentration of solid bodies
observed here does not hold for that size range, and that the disc gravity
rapidly becomes the dominating influence on the orbital evolution
of larger planetary building blocks \citep{nelson05b}.

\section{Discussion}
\label{discussion}

We have presented the results of MHD simulations of cylindrical 
accretion disc models
that include the orbital evolution of solid bodies ranging in size from 
5 centimetres to 1 metre. A two fluid approach was used for the 5 and 25 cm 
sized particles, and an N-body approach used for the 1 metre 
sized objects. Both 
methods were able to follow the radial drift of the solid bodies 
relative to the gas.

Two results emerge from these simulations: vortices 
were found to develop in the disc and to survive for the duration of the 
simulations, and dust was observed to accumulate very efficiently in 
these long--lived structures.

The first of these findings is significant as previous studies of MHD turbulence
in cylindrical discs \citep{armitage98,hawley01,steinacker&pap02}
have not reported long--lived vortices. The reason for this may be
that these structures are difficult to identify in turbulent discs,
because they tend to be weak in comparison with the transient
background features.
By--eye inspection of density plots often does not reveal 
vortices, and we found it necessary
to inspect the vorticity before being convinced that vortices were
indeed present in our models.

The issue of the formation and survival of vortices in discs is long--standing.
Although it has been shown that anticyclonic vortices can survive for 
hundreds of orbits \citep{barge&sommeria95,godon&livio99}, few mechanisms have
been put forward  to explain their formation in a barotropic fluid.
\citet{godon&livio00} suggested that small vortices resulting from 
underlying turbulence could merge to form larger, stable
anticyclonic vortices. We find that turbulence generates small scale
variations in vorticity, and it is possible that some of the
small scale regions of anticyclonic vorticity may merge to form longer
lived structures. We find that early on in the simulations,
when the disc is being relaxed prior to the dust component being added,
larger patches of underlying vorticity also exist. These are not 
apparent against the higher amplitude background fluctuations,
but become so if the magnetic field is switched off and the
turbulence is allowed to decay. The disc returns to a laminar state,
but with continued existence of the underlying vortices that are apparently
responsible for trapping dust when it is added later to the simulations.

Another possible route to their 
formation is the Papaloizou--Pringle instability \citep{pap&pringle84}.
\citet{hawley87} showed that radially--slender tori will break--up into
gaseous blobs or ``planets''. There is some evidence that the 
vortices formed in our discs occur near pressure and density maxima
caused by radial variations  in the viscous stress. It remains an open question
whether these radial density variations are susceptible to this instability,
causing the vortices we observe to form. A detailed analysis of the
origin of the vortices goes beyond the scope of this {\em letter}, and we note
that \citet{barranco&marcus05} have suggested that vortices are unstable when 
vertical stratification is included. The equation of state may also be
of some importance (see, for example, \citeauthor{klahr&bodenheimer03}
\citeyear{klahr&bodenheimer03}).We will examine these issues in a
future paper.

The second result presented in this letter is that solid bodies in the 
size range considered concentrate efficiently in regions of the disc 
where pressure maxima form due to the anticyclonic vortices described 
above. This accumulation 
is a well know process that has been studied before \citep{barge&sommeria95,
godon&livio00,johansenetal04}. The two fluid calculations demonstrated that 
the effect of trapping was more pronounced for the 30 centimetre sized objects 
than for the
5 centimetre ones because of their more rapid inward drift.
A simulation performed for 1 metre sized bodies including
the effects of the disc gravity demonstrated that gas drag forces
are very dominant in determining the evolution for these size
ranges. All simulations indicated that the trapping was very efficient,
with between 50 -- 75 percent of the solids surviving inward
migration through the full extent of the disc over runs times of
$\sim 200$ orbits. Test calculations using laminar disc models showed
complete loss of solid bodies with sizes in the range considered here.

Note that we did not include the back reaction of the solids on the gas in 
this work, which is valid only in regions where the gas density is much 
larger that the solid density. Large enhancements in the density of solids mean
that the local dynamics may become dominated by the solid bodies rather than 
gas. It will be necessary to include the back reaction of the solids
on the gas in future studies to investigate this effect. 

In the N-body approach , finite resolution effects tend to cause particles 
to clump together on the grid-scale, and it is for this reason
that we have not quoted density enhancement factors for these runs. This
issue can be addressed by adopting a sub grid scale model for the turbulence.

The effect of trapping presented in this letter 
may help overcome the problem of
very rapid migration of intermediate sized bodies during planet formation
(e.g. \citeauthor{weidenschilling77} \citeyear{weidenschilling77}).
The enhanced concentration arising from this trapping
will very likely lead to an enhanced growth rate of the solid
bodies involved {\em via} binary sticking, though the
sticking presumably occurs between
larger bodies and smaller grains that are trapped in the same region, given the
difficulty of sticking macroscopic objects together.
This will allow bodies to grow to sizes for which
the inward migration is much less rapid. 
Indeed simulations of larger
bodies in turbulent discs suggest that objects larger than 10 metres
no longer have their evolution controlled by gas drag but rather through
gravitational interaction with turbulent density fluctuations \citep{nelson05b}.

\section*{ACKNOWLEDGMENTS}
The simulations presented in this paper were performed on the 
U.K. Astrophysical Fluid Facilities (UKAFF) and on the
QMUL High Performance Computing Facility purchased under the
SRIF initiative.
The authors thank Laure Barri\`ere-Fouchet for her help in setting up 
the two fluids description used in GLOBAL. 
We also acknowledge comments provided by the referee, Hubert Klahr,
that have improved this paper.
\bibliographystyle{mn2e}
\bibliography{author}

\end{document}